\documentclass[12pt,a4paper]{article}
\usepackage{ifthen} 
\usepackage{booktabs} 
\newboolean{pdflatex}
\setboolean{pdflatex}{true} 

\newboolean{articletitles}
\setboolean{articletitles}{true} 

\newboolean{uprightparticles}
\setboolean{uprightparticles}{false} 

\newboolean{inbibliography}
\setboolean{inbibliography}{false} 

\usepackage{microtype}
\usepackage{lineno}  
\usepackage{xspace} 
\usepackage{caption} 

\usepackage{graphicx}  
\usepackage{color}
\usepackage{colortbl}
\graphicspath{{./figs/}} 

\usepackage{amsmath} 
\usepackage{amssymb}
\usepackage{amsfonts}
\usepackage{upgreek} 

\newcommand*\patchAmsMathEnvironmentForLineno[1]{%
\expandafter\let\csname old#1\expandafter\endcsname\csname #1\endcsname
\expandafter\let\csname oldend#1\expandafter\endcsname\csname
end#1\endcsname
 \renewenvironment{#1}%
   {\linenomath\csname old#1\endcsname}%
   {\csname oldend#1\endcsname\endlinenomath}%
}
\newcommand*\patchBothAmsMathEnvironmentsForLineno[1]{%
  \patchAmsMathEnvironmentForLineno{#1}%
  \patchAmsMathEnvironmentForLineno{#1*}%
}
\AtBeginDocument{%
\patchBothAmsMathEnvironmentsForLineno{equation}%
\patchBothAmsMathEnvironmentsForLineno{align}%
\patchBothAmsMathEnvironmentsForLineno{flalign}%
\patchBothAmsMathEnvironmentsForLineno{alignat}%
\patchBothAmsMathEnvironmentsForLineno{gather}%
\patchBothAmsMathEnvironmentsForLineno{multline}%
\patchBothAmsMathEnvironmentsForLineno{eqnarray}%
}

\usepackage{hyperref}    
\usepackage[all]{hypcap} 

\def\pt         {\mbox{$p_{\rm T}$}\xspace}
\newcommand{\tev}{\ifthenelse{\boolean{inbibliography}}{\ensuremath{~T\kern -0.05em eV}\xspace}{\ensuremath{\mathrm{\,Te\kern -0.1em V}}}\xspace}
\newcommand{\gev}{\ensuremath{\mathrm{\,Ge\kern -0.1em V}}\xspace}
\newcommand{\stat}{\ensuremath{\mathrm{\,(stat)}}\xspace}
\newcommand{\syst}{\ensuremath{\mathrm{\,(syst)}}\xspace}
\def\fb   {\ensuremath{\mbox{\,fb}}\xspace}

\def\lhcb {\mbox{LHCb}\xspace}
 \def\Pb      {\ensuremath{\mathrm{b}}\xspace}                 
 \def\Pc      {\ensuremath{\mathrm{c}}\xspace} 
\def\bquark    {{\ensuremath{\Pb}}\xspace}
\def\cquark    {{\ensuremath{\Pc}}\xspace}
\def\pythia     {\mbox{\textsc{Pythia}}\xspace}
\def\evtgen     {\mbox{\textsc{EvtGen}}\xspace}
\def\photos     {\mbox{\textsc{Photos}}\xspace}
\def\geant      {\mbox{\textsc{Geant4}}\xspace}

\usepackage{cite} 
\usepackage{mciteplus}

\usepackage{longtable} 
\usepackage{chngpage}

\def \light {\ensuremath{\rm light\mbox{-}parton}\xspace}

\def \muptr {\ensuremath{\pt(\mu)/\pt(j_\mu)}\xspace}
\def \wc {\ensuremath{W\!+\!c}\xspace}
\def \wb {\ensuremath{W\!+\!b}\xspace}

\begin{document}

\renewcommand{\thefootnote}{\fnsymbol{footnote}}
\setcounter{footnote}{1}

\begin{titlepage}
\pagenumbering{roman}

\vspace*{-1.5cm}
\centerline{\large EUROPEAN ORGANIZATION FOR NUCLEAR RESEARCH (CERN)}
\vspace*{1.5cm}
\hspace*{-0.5cm}
\begin{tabular*}{\linewidth}{lc@{\extracolsep{\fill}}r}
\ifthenelse{\boolean{pdflatex}}
{\vspace*{-2.7cm}\mbox{\!\!\!\includegraphics[width=.14\textwidth]{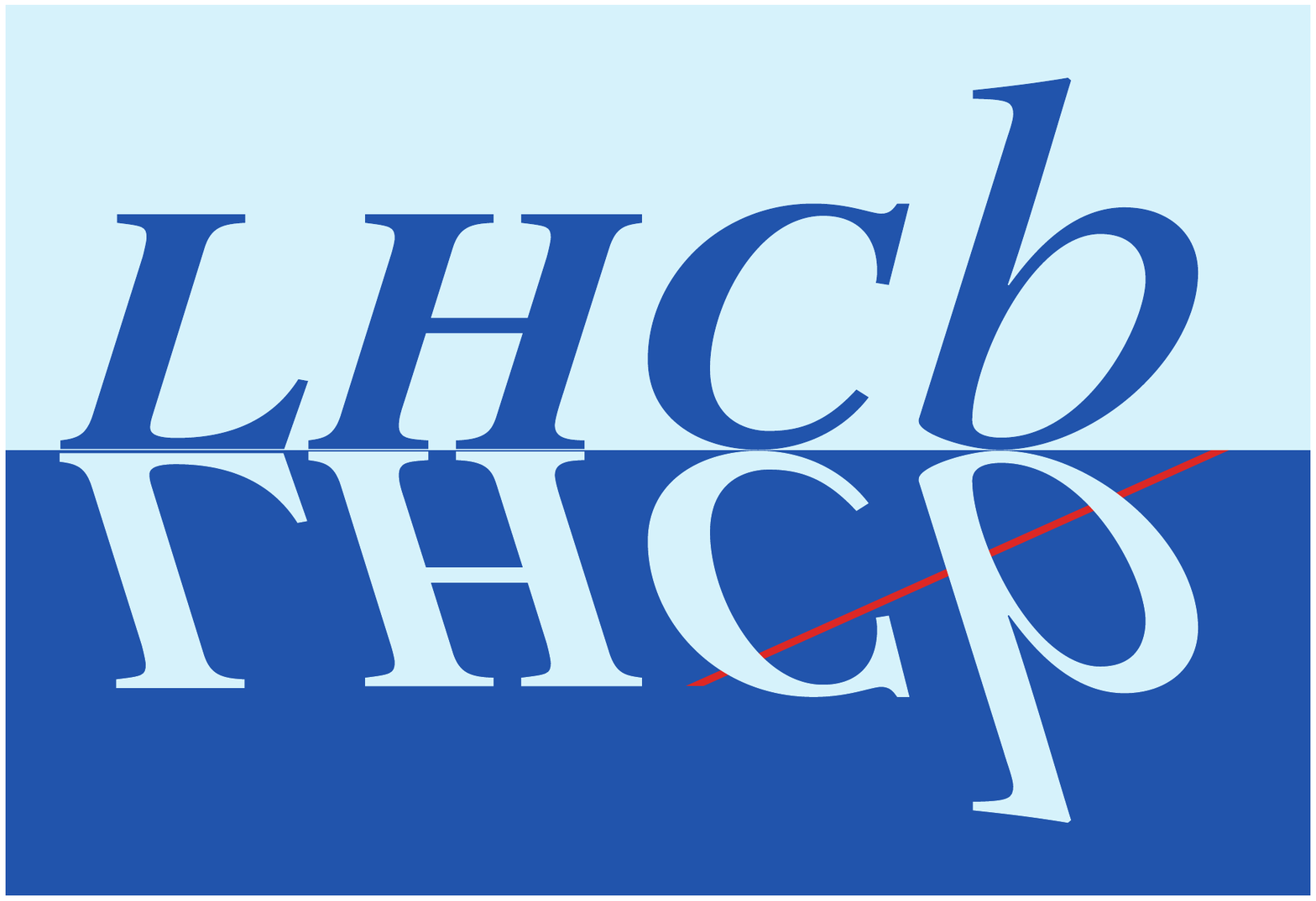}} & &}%
{\vspace*{-1.2cm}\mbox{\!\!\!\includegraphics[width=.12\textwidth]{figs/lhcb-logo.eps}} & &}%
\\
 & & CERN-PH-EP-2015-132 \\  
 & & LHCb-PAPER-2015-022 \\  
 & & September 8, 2015 \\ 
 & & \\
\end{tabular*}

\vspace*{4.0cm}

{\bf\boldmath\huge
\begin{center}
  First observation of top quark production in the forward region
\end{center}
}

\vspace*{0.5cm}

\begin{center}
The LHCb collaboration\footnote{Authors are listed at the end of this Letter.}
\end{center}

\vspace{\fill}

\begin{abstract}
  \noindent
Top quark production in the forward region in proton-proton collisions is observed for the first time.
The $W\!+\!b$ final state with $W\to\mu\nu$ is reconstructed using  muons with a transverse momentum, \pt, larger than 25\gev in the pseudorapidity range $2.0<\eta<4.5$.  
The $b$ jets are required to have $50 < \pt < 100\gev$ and $2.2 < \eta < 4.2$, while the transverse component of
the sum of the muon and $b$-jet momenta must satisfy $\pt > 20\gev$. 
The results are based on data corresponding to integrated luminosities of 1.0 and $2.0\fb^{-1}$ collected at center-of-mass energies of 7 and 8\tev by LHCb.  The inclusive top quark production cross-sections in the fiducial region are
\begin{eqnarray}
\sigma({\rm top})[7\tev] &=& \!239\pm53\stat\pm33\syst\pm24\,({\rm theory})\fb\,,  \nonumber \\
\sigma({\rm top})[8\tev] &=& \!289\pm43\stat\pm40\syst\pm29\,({\rm theory})\fb\,. \nonumber 
\end{eqnarray}
These results, along with the observed differential yields and charge asymmetries, are in agreement with next-to-leading order Standard Model predictions.
\end{abstract}

\vspace*{0.5cm}

\begin{center}
  Phys.\ Rev.\ Lett.\ {\bf 115} (2015) 112001
\end{center}

\vspace{\fill}

{\footnotesize 
\centerline{\copyright~CERN on behalf of the \lhcb collaboration, license \href{http://creativecommons.org/licenses/by/4.0/}{CC-BY-4.0}.}}
\vspace*{2mm}

\end{titlepage}

\newpage
\setcounter{page}{2}
\mbox{~}
\newpage

\renewcommand{\thefootnote}{\arabic{footnote}}
\setcounter{footnote}{0}

\pagestyle{plain} 
\setcounter{page}{1}
\pagenumbering{arabic}


\clearpage

The production of top quarks ($t$) from  proton-proton ($pp$) collisions in the forward region is of considerable experimental and theoretical interest. 
In the Standard Model (SM), four processes make significant contributions to top quark production: $t\bar{t}$ pair production; 
single top production via processes mediated by a $W$ boson in the $t$-channel ($qb \to q^{\prime}t$) or in the 
$s$-channel ($q\bar{q}^{\prime} \to t\bar{b}$); 
and single top produced in association with a $W$ boson ($gb \to tW$). 
The initial-state $b$ quarks arise from gluon splitting to $b\bar{b}$ pairs or from the intrinsic $b$ quark content in the proton. 
Top quarks decay almost entirely via $t\to Wb$. 
The SM predicts that about 75\% of $t\to Wb$ decays in the forward region 
are due to $t\bar{t}$ pair production.  The remaining 25\% are mostly due to $t$-channel single-top production, with $s$-channel and associated single-top production making percent-level contributions.

The enhancement at forward rapidities of $t\bar{t}$ production via $q\bar{q}$ and $qg$ scattering, relative to $gg$ fusion,
 can result in larger charge asymmetries, which may be sensitive to physics beyond the SM~\cite{PhysRevLett.107.082003,gauld}. Forward $t\bar{t}$ events can be used to constrain the gluon parton distribution function (PDF) at large momentum fraction, resulting in reduced theoretical uncertainty for many SM predictions~\cite{gauld2}. Furthermore, 
both single-top and $t\bar{t}$ cross-section measurements in the forward region will provide important experimental tests of differential next-to-next-to-leading order theoretical calculations as they become available~\cite{Czakon:2014xsa}.

This Letter reports the first observation of top quark production in the forward region. The data used correspond to integrated luminosities of 1.0 and $2.0\fb^{-1}$ collected at center-of-mass energies of $\sqrt{s}=7$ and 8\tev in $pp$ collisions with the LHCb detector. 
The $W$ bosons are reconstructed using the $W\to\mu\nu$ decay with  muons having a transverse momentum, \pt, larger than 25\gev ($c=1$ throughout this Letter) in the pseudorapidity range $2.0<\eta<4.5$.  
The analysis is performed using jets clustered with the anti-$k_{\rm T}$ algorithm~\cite{1126-6708-2008-04-063} using a distance parameter $R=0.5$.
The jets are required to have $50 < \pt < 100\gev$ and $2.2 < \eta < 4.2$. 
The muon and jet ($j$) must be separated by $\Delta R(\mu,j) > 0.5$, with $\Delta R \equiv \sqrt{\Delta\eta^2 + \Delta\phi^2}$. Here $\Delta\eta\,(\Delta\phi)$ is the difference in pseudorapidity (azimuthal angle) between the muon and jet momenta.
The transverse component of the sum of the muon and jet momenta must satisfy $\pt(\mu+j)  \equiv \left(\vec{p}(\mu) + \vec{p}(j)\right)_{\rm T} > 20\gev$.

The \lhcb detector is a single-arm forward spectrometer covering the pseudorapidity range $2<\eta <5$, designed for the study of particles containing \bquark or \cquark quarks.  It is described in detail in Refs.~\cite{Alves:2008zz,Aaij:2014jba}.  
The trigger~\cite{LHCb-DP-2012-004} consists of a hardware stage, based on information from the calorimeter and muon systems, followed by a software stage, which applies a full event reconstruction. This analysis requires at least one muon candidate that satisfies the trigger requirement of $\pt>10\gev$. Global event cuts (GECs), which prevent high-occupancy events from dominating the processing time of the software trigger, have an efficiency of about $90\%$ for $W+$jet and top quark events.

Simulated $pp$ collisions
are generated using \pythia~\cite{Sjostrand:2006za,*Sjostrand:2007gs} with an \lhcb configuration~\cite{LHCb-PROC-2010-056}. Decays of hadronic particles are described by \evtgen~\cite{Lange:2001uf} in which final-state radiation is generated using \photos~\cite{Golonka:2005pn}. The interaction of the generated particles with the detector, and its response, are implemented using the \geant toolkit~\cite{Allison:2006ve, *Agostinelli:2002hh} as described in Ref.\cite{LHCb-PROC-2011-006}. Further theory calculations are performed at next-to-leading order (NLO) with MCFM~\cite{Campbell:2000bg} and the CT10 PDF set~\cite{Lai:2010vv}, and are cross-checked using {\sc PowhegBox}~\cite{Alioli:2010xd} with hadronization simulated by {\sc Pythia}.
The theoretical uncertainty on the cross-section predictions is a combination of PDF, scale, and strong-coupling ($\alpha_s$) uncertainties.
The PDF and scale uncertainties are evaluated following Refs.~\cite{Lai:2010vv} and \cite{Hamilton:2013fea}, respectively.
The $\alpha_s$ uncertainty is evaluated as the envelope obtained using $\alpha_s(M_Z) \in [0.117, 0.118, 0.119]$ in the theory calculations.
 
The event selection is the same as that in Ref.\cite{LHCb-PAPER-2015-021} but a reduced fiducial region is used to enhance the top quark contribution relative to direct \wb production.
The signature for $W+$jet events is an isolated high-\pt muon and a well-separated jet originating from the same $pp$ interaction. Signal events are selected by requiring a high-\pt muon candidate and at least one jet with $\Delta R(\mu,j) > 0.5$. For each event, the highest-\pt muon candidate that satisfies the trigger requirements is selected, along with the highest-\pt jet from the same $pp$ collision. 
The primary background to top quark production is direct \wb production; however, $Z\!+\!b$ events, with one muon undetected in the decay $Z\to\mu\mu$, and di-$b$-jet events also contribute to the $\mu\!+\!b$-jet final state.

The anti-$k_T$ clustering algorithm is used as implemented in
\textsc{FastJet}~\cite{fastjet}.
Information from all the detector
subsystems is used to create charged and neutral particle inputs to
the jet-clustering algorithm using a particle flow approach~\cite{LHCb-PAPER-2013-058}. 
The reconstructed jets must fall within the pseudorapidity range $2.2 < \eta(j) < 4.2$. The reduced $\eta(j)$ acceptance ensures nearly uniform jet reconstruction and heavy-flavor tagging efficiencies.
The momentum of a reconstructed jet is corrected to obtain an unbiased
estimate of the true jet momentum. The correction factor, typically between
0.9 and 1.1, is determined from simulation and depends on the jet \pt and $\eta$, the fraction of the jet \pt measured with the
tracking system, and the number of $pp$ interactions in the event.

The high-\pt muon candidate is not removed from the anti-$k_T$ inputs and so is clustered into a jet.
 This jet, referred to as the muon jet and denoted as $j_{\mu}$,
is used to discriminate between $W+$jet and dijet events~\cite{LHCb-PAPER-2015-021}.
No correction is applied to the momentum of the muon jet.
The requirement $\pt(j_\mu+j) > 20\gev$ is made to suppress dijet backgrounds, which are well balanced in \pt, unlike $W+$jet events, where there is undetected energy from the neutrino.  
Events with a second, oppositely charged, high-\pt muon candidate from the same $pp$ collision are vetoed.   However, when the dimuon invariant mass is in the range $60 < M(\mu^+\mu^-) < 120\gev$, such events are selected as $Z(\mu\mu)+$jet candidates, which are used to determine the $Z+$jet background. 

The jets are identified (tagged) as originating from the hadronization of a $b$ or $c$ quark by the presence of a secondary vertex (SV) with $\Delta R < 0.5$ between the jet axis and the SV direction of flight, defined by the vector from the $pp$ interaction point to the SV position.  
Two boosted decision trees (BDTs)~\cite{Breiman,Adaboost}, 
trained on the characteristics of the SV and the jet, are used to 
separate heavy-flavor jets from \light jets, 
and to separate $b$ jets from $c$ jets.  
The two-dimensional distribution of the BDT responses observed in data is fitted to obtain the SV-tagged $b$, $c$ and \light jet yields. The SV-tagger algorithm is described in Ref.~\cite{LHCb-PAPER-2015-016}, where the heavy-flavor tagging efficiencies and \light mistag probabilities are measured in data.
The data samples used in Ref.~\cite{LHCb-PAPER-2015-016} are too small
 to validate the performance of the SV-tagger algorithm in the $\pt(j) > 100\gev$ region. Furthermore, the mistag probability of \light jets increases with jet \pt. Therefore, only jets with $\pt < 100\gev$ are considered in the fiducial region, which according to simulation retains about 80\% of all top quark events.

Inclusive $W+$jet production, {\em i.e.}\ where no SV-tag requirement is made on the jet, is only contaminated at the percent level by processes other than direct $W+$jet production. Therefore, $W+$jet production is used to validate both the theory predictions and the modeling of the detector response.
Furthermore, the SM prediction for $\sigma(Wb)/\sigma(Wj)$ has a smaller relative uncertainty than $\sigma(Wb)$ alone, since the theory uncertainties partially cancel in the ratio.
The analysis strategy is to first measure the $W+$jet yields, and then to obtain predictions for the yields of direct \wb production using the prediction for $\sigma(Wb)/\sigma(Wj)$.  To an excellent approximation, many experimental effects, {\em e.g.}\ the muon reconstruction efficiency, are expected to be the same for both samples and do not need to be considered in the direct \wb yield prediction.

\begin{figure}
  \begin{center}
    \includegraphics[width=0.49\textwidth]{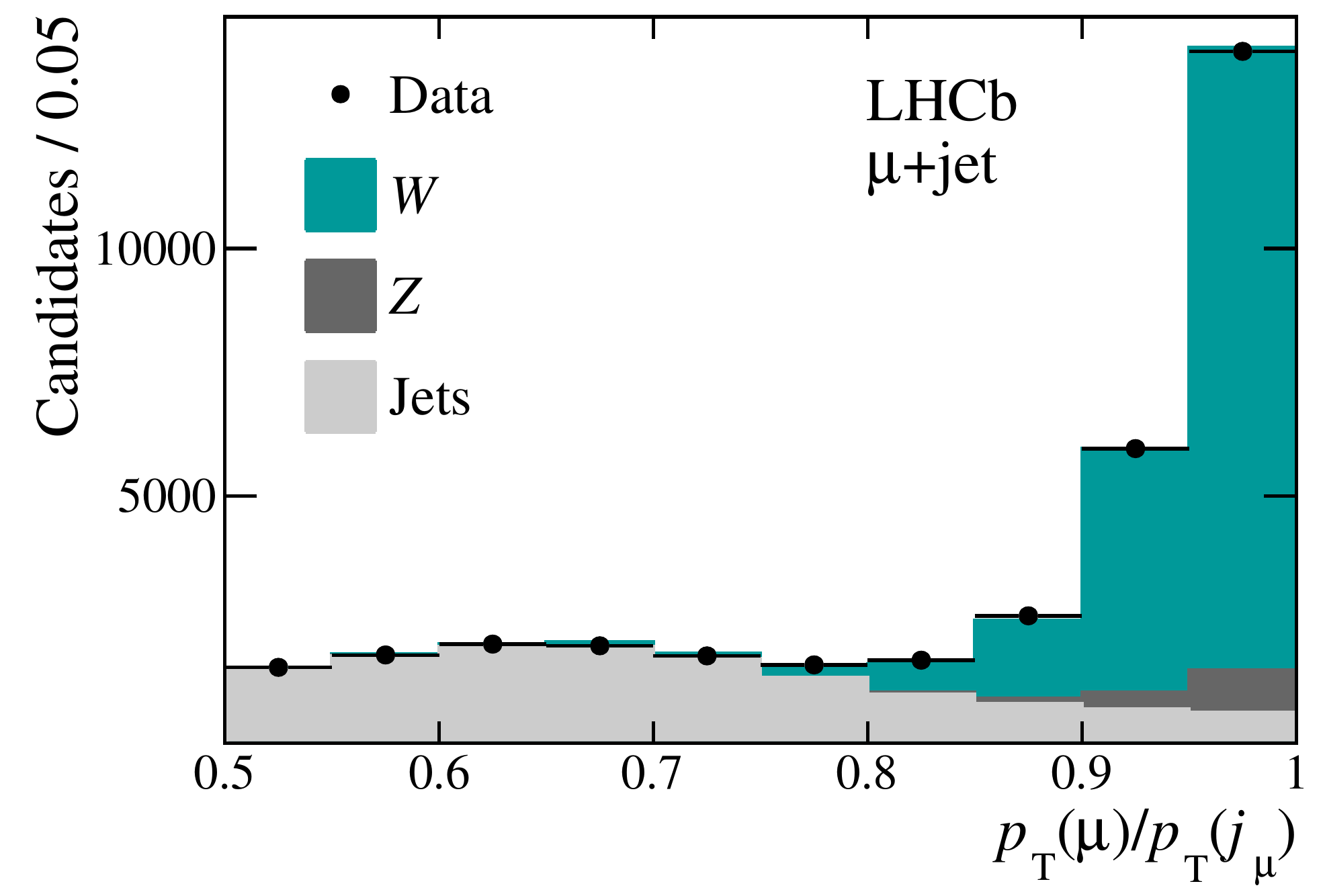}
    \caption{Distribution of \muptr with fit overlaid for all $W+$jet candidates.  
\label{fig:wj_muptr}}
  \end{center}
\end{figure}

\begin{figure}
  \begin{center}
    \includegraphics[width=0.49\textwidth]{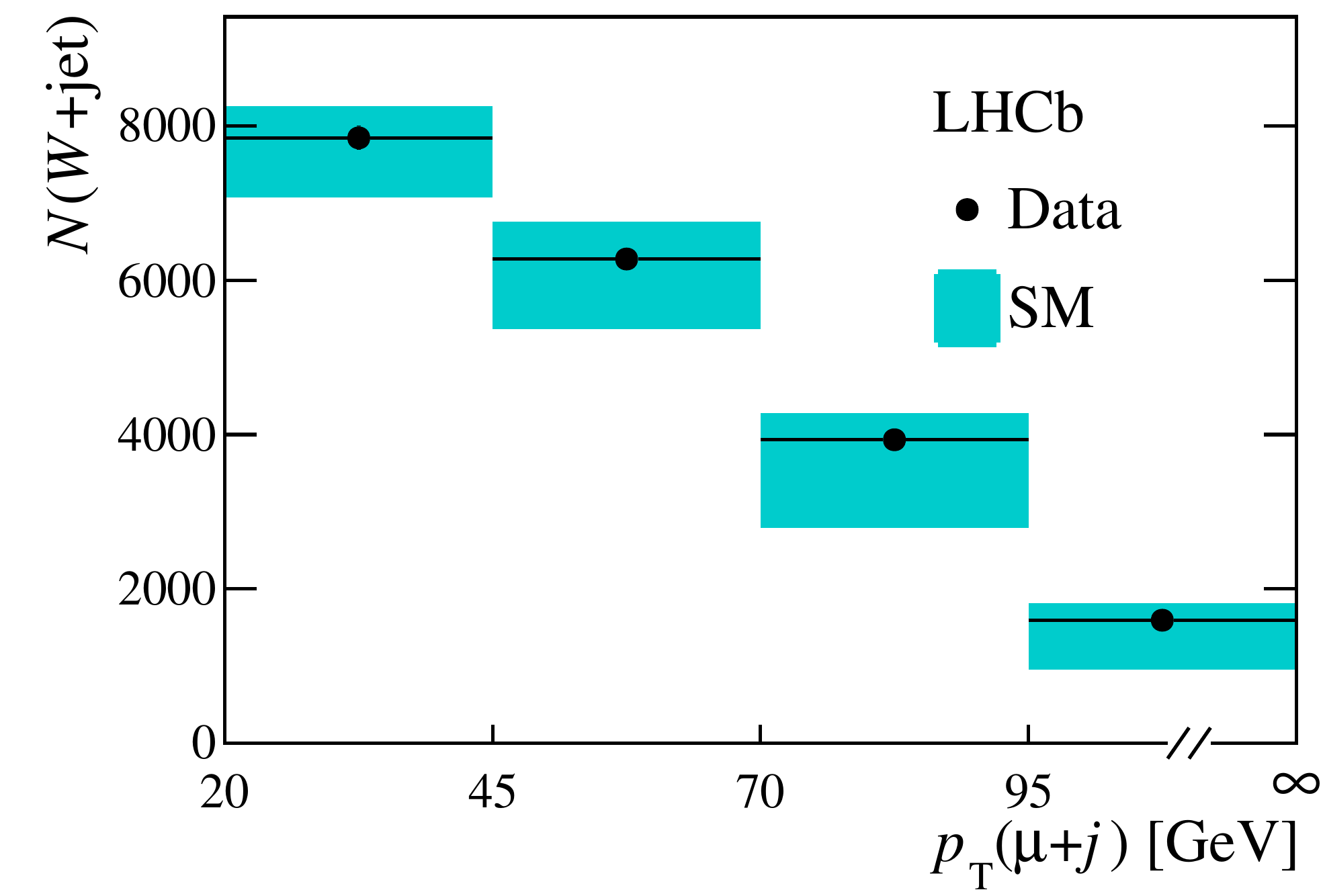}
    \includegraphics[width=0.49\textwidth]{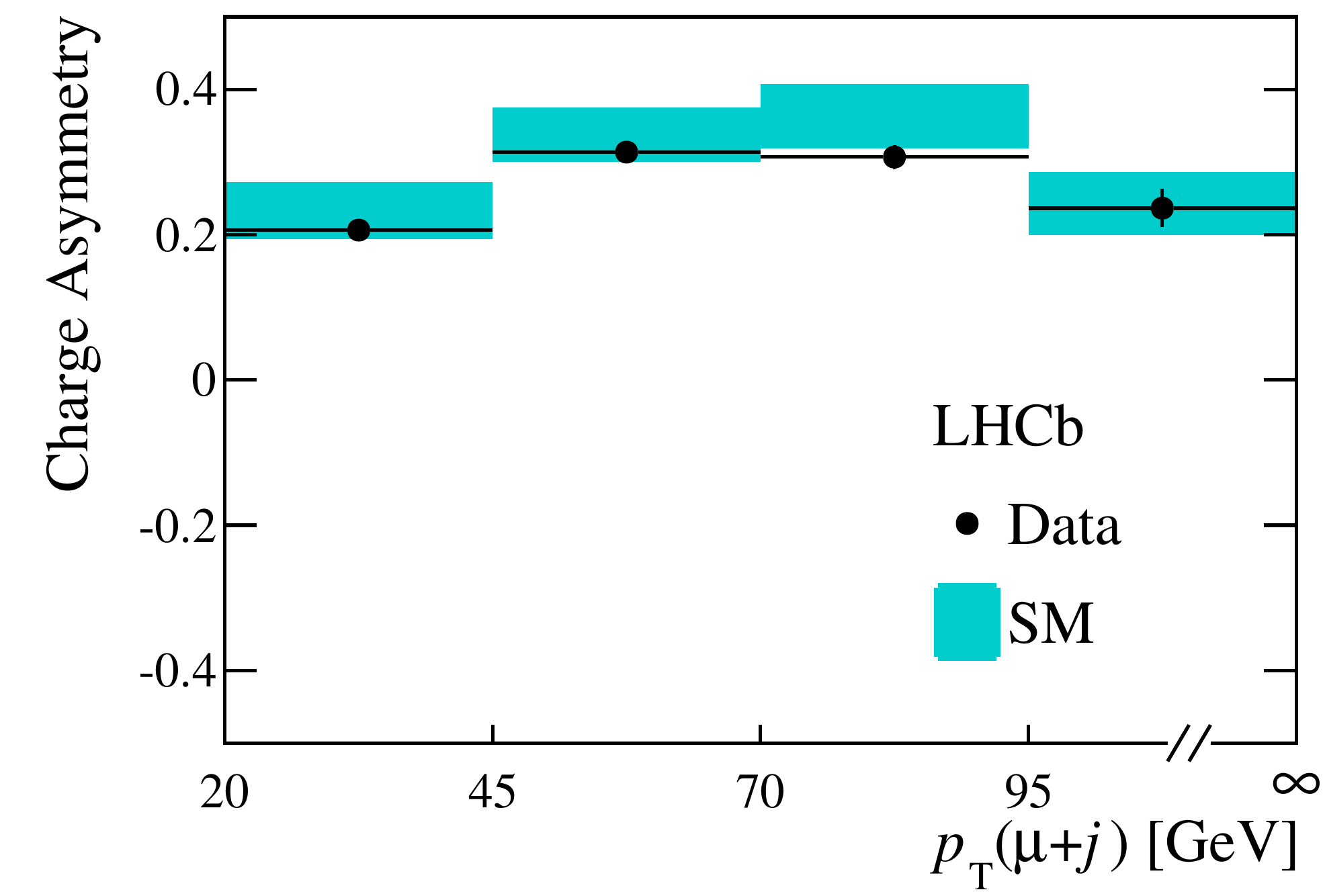}
    \caption{Results for the inclusive $W+$jet yield (left) and charge asymmetry (right) versus $\pt(\mu+j)$ compared to SM predictions at NLO obtained using MCFM.  The data error bars are smaller than the marker size,  
the SM uncertainties are highly correlated across $\pt(\mu+j)$ bins.
\label{fig:wj_results}}
  \end{center}
\end{figure}

The $W+$jet yield is determined by performing a fit to the $\pt(\mu)/\pt(j_{\mu})$ distribution with templates, histograms obtained from data, as described in Ref.~\cite{LHCb-PAPER-2015-021}. 
The $Z+$jet contribution is fixed from the fully reconstructed $Z(\mu\mu)+$jet yield, where the probability for one of the muons to escape detection is obtained using simulation.  
The contributions of $b$, $c$, and light-parton jets are each free to vary in the fit.  
Figure~\ref{fig:wj_muptr} shows the fit for all candidates in the data sample. Such a fit is performed for each muon charge separately in bins of $\pt(\mu+j)$;  the differential $W+$jet yield and charge asymmetry, defined as $[\sigma(W^+j)-\sigma(W^-j)]/[\sigma(W^+j)+\sigma(W^-j)]$,
 are given in Fig.~\ref{fig:wj_results}. 

To compare the data to theory predictions, the detector response must be taken into account. All significant aspects of the detector response are determined using data-driven techniques. The muon trigger, reconstruction, and selection efficiencies are determined using $Z\to\mu\mu$ events~\cite{LHCb-PAPER-2015-001,LHCb-PAPER-2013-058}. 
The GEC efficiency is obtained following Ref.~\cite{LHCb-PAPER-2013-058}: an alternative dimuon trigger requirement, which requires a looser GEC, is used to determine the fraction of events that are rejected.
Contamination from $W\to\tau\to\mu$ decays is estimated to be 2.5\% using both simulated $W+$jet events and inclusive $W$ data samples~\cite{LHCb-PAPER-2014-033}.  The fraction of muons that migrate out of the fiducial region due to final-state radiation is about 1.5\%~\cite{LHCb-PAPER-2014-033}. 

Migration of events in jet \pt due to the detector response is studied with
 a data sample enriched in $b$ jets using SV tagging. The $\pt({\rm SV})/\pt(j)$ distribution observed in data is compared to templates obtained from simulation in bins of jet \pt. The resolution and scale for each jet \pt bin are varied in simulation  to find the best description of the data and to construct a detector response matrix.
Figure~\ref{fig:wj_results} shows that the SM predictions, obtained with all detector response effects applied, agree with the inclusive $W+$jet data.

The yields of \wc and \wb, which includes $t \to Wb$ decays, are determined using the subset of candidates with an SV-tagged jet and binned according to \muptr.  In each \muptr bin, the two-dimensional SV-tagger BDT-response distributions are fitted to determine the yields of $c$-tagged and $b$-tagged jets, which are used to form the \muptr distributions for candidates with $c$-tagged and $b$-tagged jets.  
These \muptr distributions are fitted to determine the SV-tagged \wc and \wb yields.

\begin{figure}
  \begin{center}
    \includegraphics[width=0.49\textwidth]{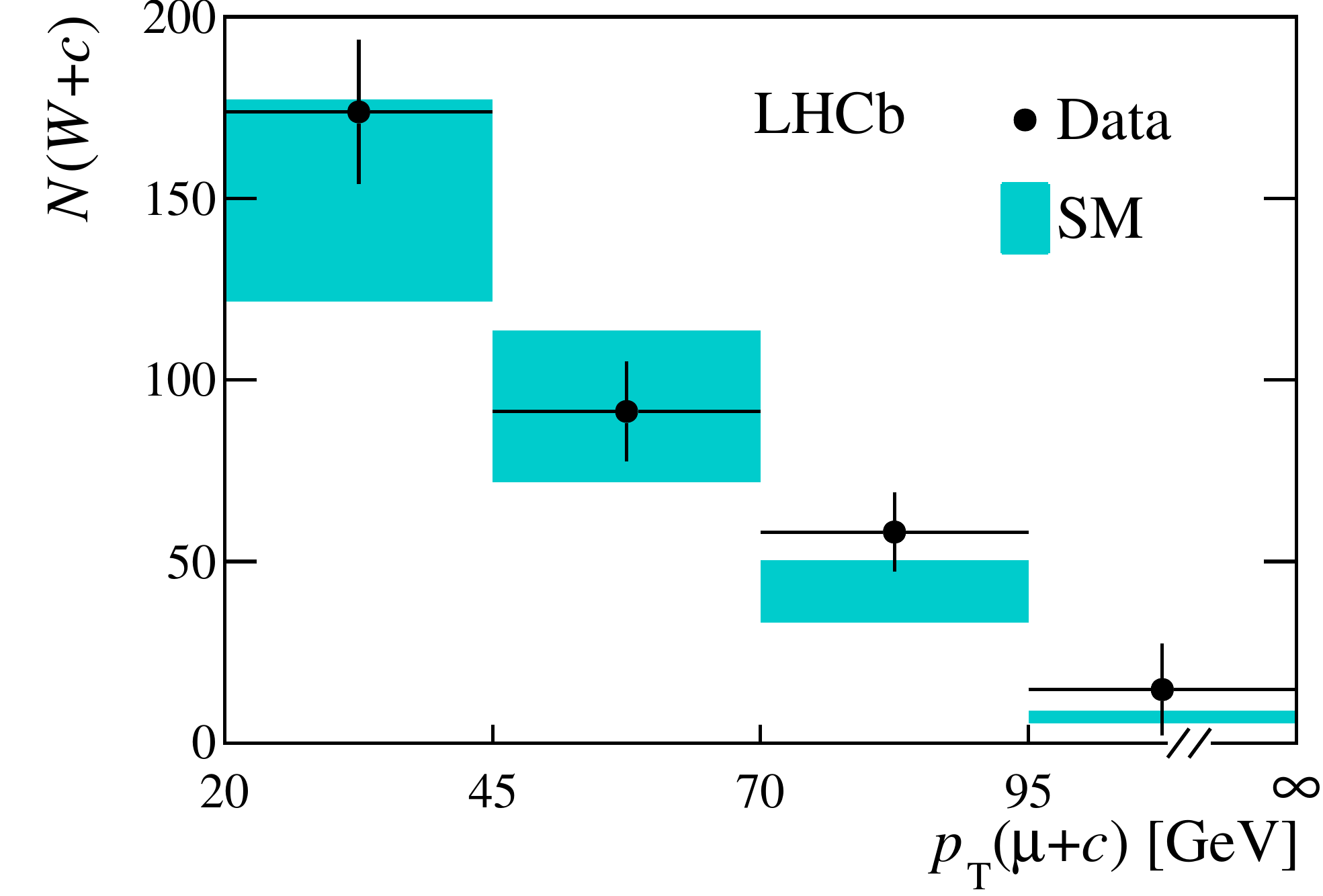}
    \caption{Results for \wc compared to SM predictions at NLO obtained using MCFM.\label{fig:wc_results}}
  \end{center}
\end{figure}

A fit to the $\pt(\mu)/\pt(j_\mu)$ distribution built from the $c$-tagged jets from the full data sample is provided as supplemental material to this Letter~\cite{Supp}.
Figure~\ref{fig:wc_results} shows that the \wc yield versus $\pt(\mu+c)$ 
agrees with the SM prediction. 
Since the \wc final state does not have any significant contributions from diboson or top quark production in the SM, this comparison validates the analysis procedures.

\begin{figure}
  \begin{center}
    \includegraphics[width=0.49\textwidth]{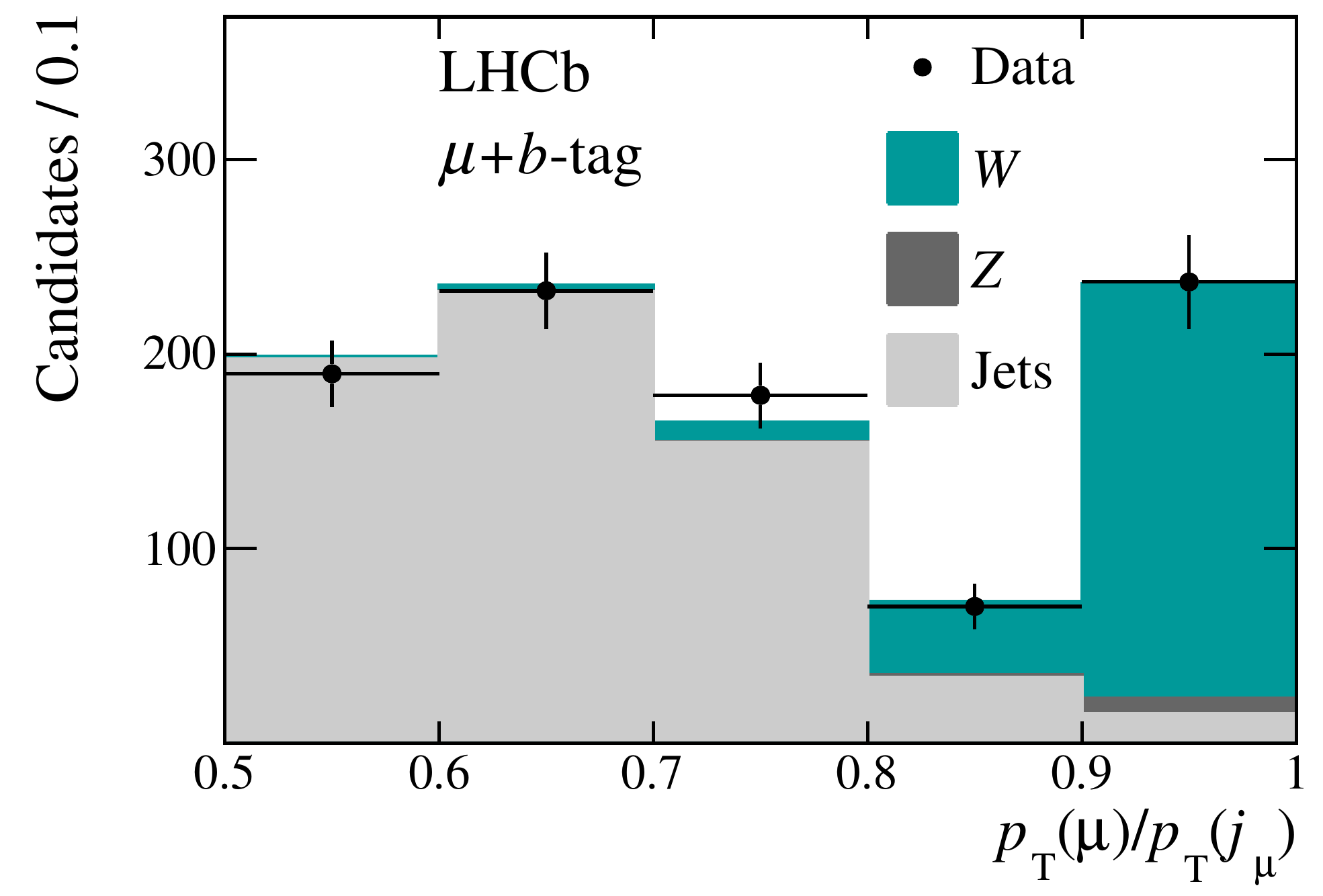}
    \caption{Distribution of \muptr with fit overlaid for all \wb candidates.
\label{fig:wb_muptr}}
  \end{center}
\end{figure}

\begin{figure}
  \begin{center}
    \includegraphics[width=0.49\textwidth]{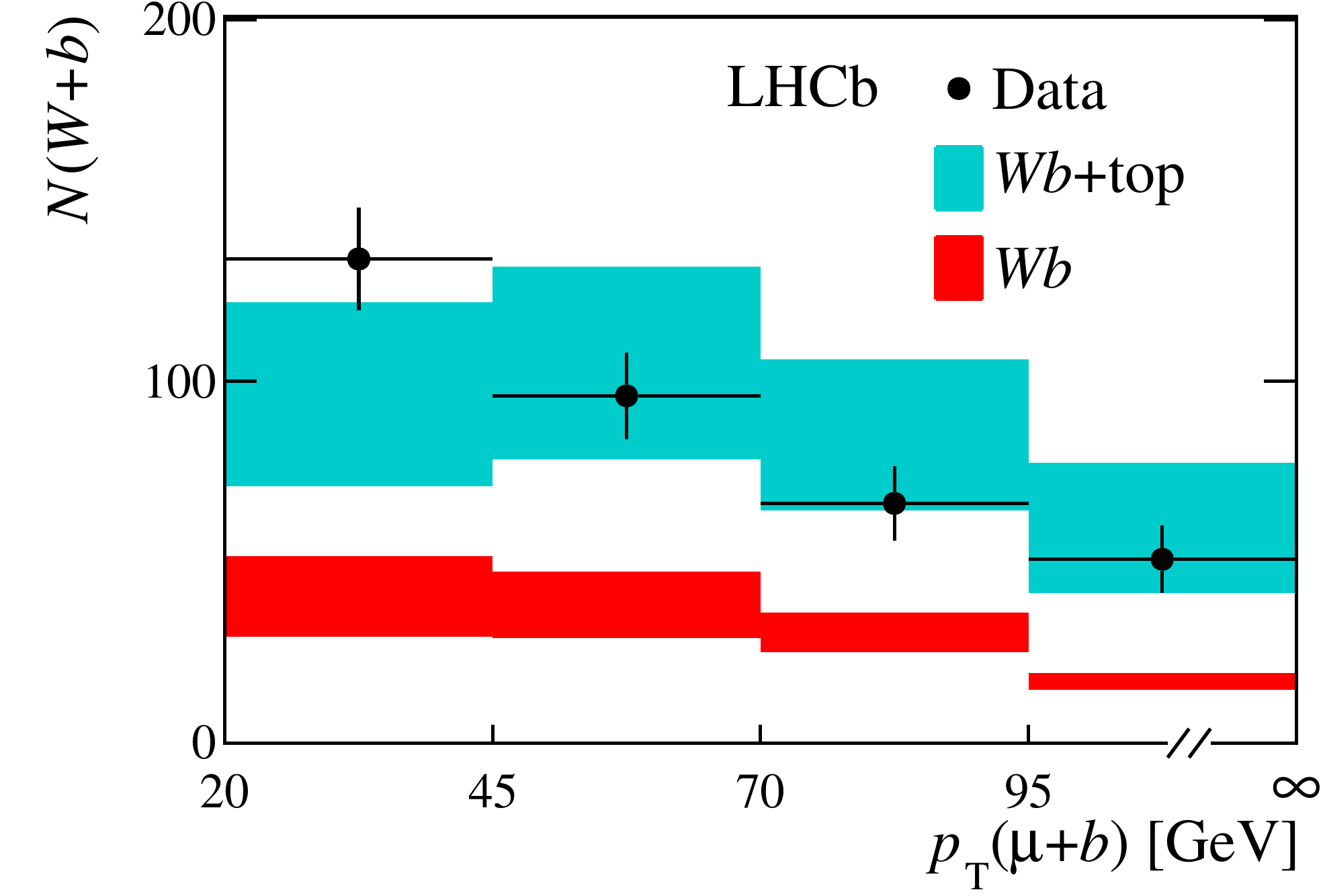}
    \includegraphics[width=0.49\textwidth]{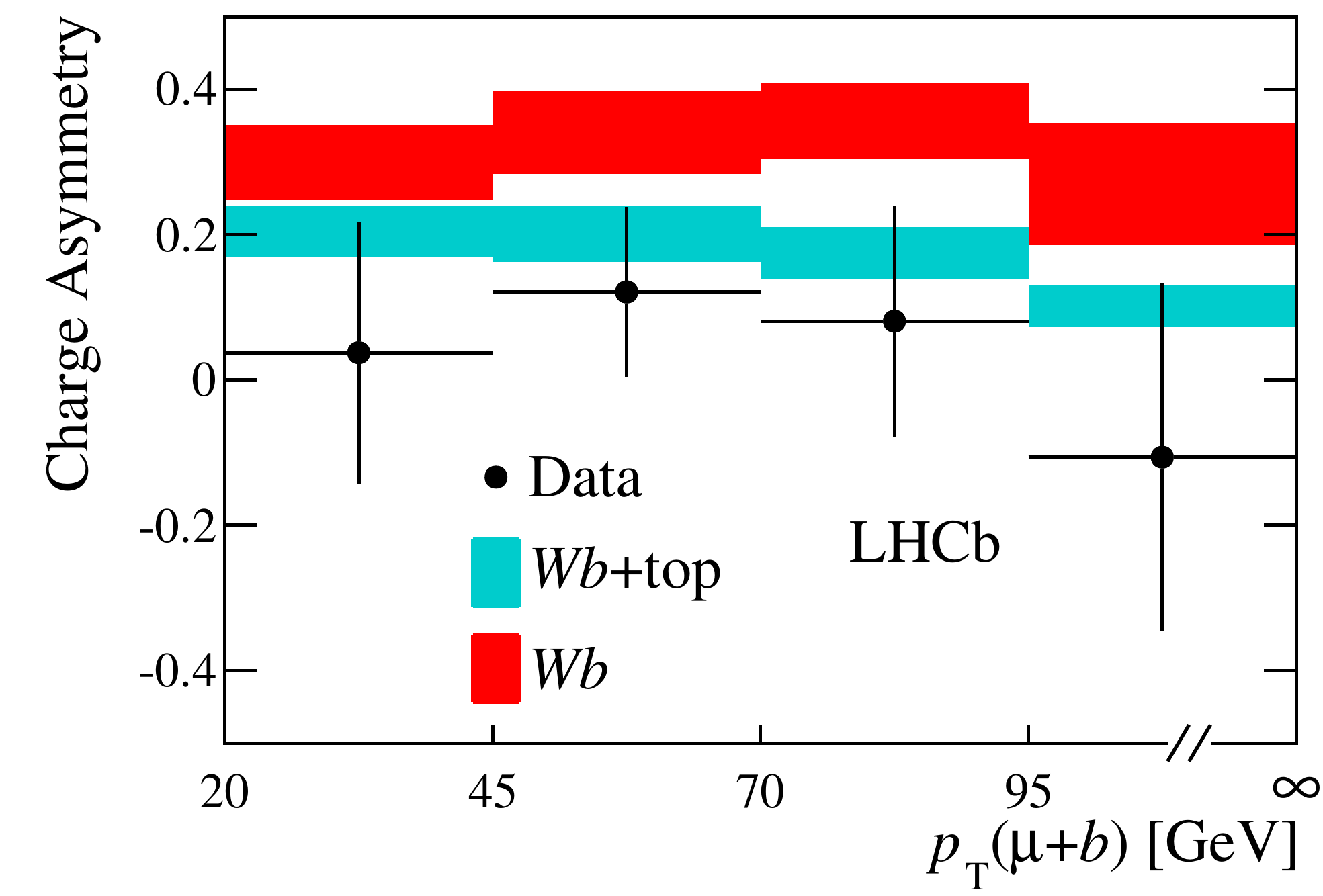}
    \caption{Results for the \wb yield (left) and charge asymmetry (right) versus $\pt(\mu+b)$ compared to SM predictions obtained at NLO using MCFM.\label{fig:wb_results}}
  \end{center}
\end{figure}

Figure~\ref{fig:wb_muptr} shows a fit to the $\pt(\mu)/\pt(j_\mu)$ distribution built from the $b$-tagged jets from the full data sample. 
For $\pt(\mu)/\pt(j_\mu) > 0.9$ the data are dominantly from $W$ decays.
Figure~\ref{fig:wb_results} shows the yield and charge asymmetry distributions obtained as a function of $\pt(\mu+b)$. 
The direct \wb prediction is determined by scaling the inclusive $W+$jet distribution observed in data by the SM prediction for $\sigma(Wb)/\sigma(Wj)$ and by the $b$-tagging efficiency measured in data~\cite{LHCb-PAPER-2015-016}. 
As can be seen, the data cannot be described by the expected direct \wb contribution alone.
The observed yield is about three times larger than the SM prediction without a top quark contribution, while the SM prediction
including both $t\bar{t}$ and single-top production does describe the data well.  

In Ref.~\cite{LHCb-PAPER-2015-021}, $W\!+b$ is studied in a larger fiducial region ($\pt(\mu) > 20\gev, \pt(j) > 20\gev$), where the top quark contribution is expected to be about half as large as that of direct $W\!+b$ production. 
The ratio $[\sigma(Wb)\!+\!\sigma({\rm top})]/\sigma(Wj)$ is measured in the larger fiducial region to be $1.17\pm0.13\stat\pm0.18\syst$\% at $\sqrt{s}=7$\tev and $1.29\pm0.08\stat\pm0.19\syst$\% at $\sqrt{s}=8$\tev.  These results agree with SM predictions, that include top quark production, of $1.23\pm0.24\%$ and $1.38\pm0.26\%$, respectively.  
This validates the direct \wb prediction, since direct \wb production is the dominant contribution to the larger fiducial region. 

Various sources of systematic uncertainties are considered and summarized in Table~\ref{tab:sys}. 
The direct \wb prediction is normalized using the observed inclusive $W+$jet data yields. Therefore, most experimental systematic uncertainties cancel to a good approximation.  

Since the muon kinematic distributions in $W+$jet and \wb are similar, all muon-based uncertainties are negligible with the exception of the trigger GEC efficiency.  
 The data-driven GEC study discussed above shows that the efficiencies are consistent for $W+$jet and \wb, with the statistical precision of this study assigned as the systematic uncertainty. 
Mismodeling of the \muptr distributions largely cancels, since this shifts the inclusive $W+$jet and \wb final-state yields by the same amount, leaving the observed excess over the expected direct \wb yield unaffected.  The one exception is possible mismodeling of the dijet templates, since the flavor content of the dijet background is not the same in the two samples.  Variations of these templates are considered 
and a relative uncertainty of 5\% is assigned on the $W$ boson yields.

The jet reconstruction efficiencies for heavy-flavor and light-parton jets in simulation are found to be consistent within 2\%, which is assigned as the systematic uncertainty for flavor-dependencies in the jet-reconstruction efficiency.  
The SV-tagger BDT templates used in this analysis are two-dimensional histograms obtained from the data samples enriched in $b$ and $c$ jets used in Ref.~\cite{LHCb-PAPER-2015-016}.  Following 
Refs.~\cite{LHCb-PAPER-2015-016,LHCb-PAPER-2015-021}, a 5\% uncertainty on the $b$-tagged yields is assigned due to uncertainty in these templates.  
The precision of the $b$-tagging efficiency measurement (10\%) in data~\cite{LHCb-PAPER-2015-016} is assigned as an additional uncertainty.

\begin{table}
  \begin{center}
    \caption{\label{tab:sys} Relative systematic uncertainties. The symbol $\dagger$ denotes an uncertainty that only applies to the cross-section measurement and not the significance determination. Only the luminosity uncertainty depends on $\sqrt{s}$: 2\% at 7~TeV and 1\% at 8~TeV.}
    \newcommand{\pad}{\phantom{$^\dagger$}}
    \begin{tabular}{lr}
      \toprule
      source & uncertainty \\
      \midrule
      GEC &  2\%\pad \\
      \muptr templates & 5\%\pad \\
      jet reconstruction & 2\%\pad \\
      SV-tag BDT templates & 5\%\pad \\
      $b$-tag efficiency & 10\%\pad \\
      trigger \& $\mu$ selection & $2\%^{\dagger}$ \\
      jet energy & $5\%^{\dagger}$ \\
      $W\to\tau\to\mu$ & $1\%^{\dagger}$ \\
      luminosity & 1--2\%$^{\dagger}$ \\
      \midrule
      Total & 14\% \\
      \midrule
      Theory & 10\% \\
      \bottomrule
    \end{tabular}
  \end{center}
\end{table}

To determine the statistical significance of the top quark contribution, a binned profile likelihood test is performed.
The top quark distribution and charge asymmetry versus $\pt(\mu+b)$ are obtained from the SM predictions.  
The total top quark yield is allowed to vary freely.
 Systematic uncertainties, both theoretical and experimental, are handled as Gaussian constraints. The profile likelihood technique is used to compare the SM hypotheses with and without a top quark contribution.
 The significance obtained using Wilks theorem~\cite{Wilks:1938dza} is $5.4\sigma$, confirming the observation of top quark production in the forward region.

The yield and charge asymmetry distributions versus $\pt(\mu+b)$ observed at $\sqrt{s}=7$ and 8\tev are each consistent with the SM predictions.
The excess of the observed yield relative to the direct \wb prediction at each $\sqrt{s}$ is attributed to top quark production, and used to measure the cross-sections.
Some additional systematic uncertainties that apply to the cross-section measurements do not factor into the significance determination.
The uncertainties due to the muon trigger, reconstruction, and selection efficiencies are taken from the data-driven studies of Refs.~\cite{LHCb-PAPER-2015-001,LHCb-PAPER-2013-058}.  
The uncertainty due to the jet energy determination is obtained from the data-driven study used to obtain the detector response matrix.
The uncertainty due to $W\to\tau\to\mu$ contamination is taken as the difference between the contamination in simulation versus that of a data-driven study of  inclusive $W\to\mu\nu$ production~\cite{LHCb-PAPER-2014-033}. 
The luminosity uncertainty is described in detail in Ref.~\cite{LHCb-PAPER-2014-047}.
The total systematic uncertainty is obtained by adding the individual contributions in quadrature.

The resulting inclusive top production cross-sections 
in the fiducial region defined by 
 $\pt(\mu) > 25\gev$, $2.0 < \eta(\mu) < 4.5$, ${50 < \pt(b) < 100\gev}$, $2.2 < \eta(b) < 4.2$, $\Delta R(\mu,b) > 0.5$, and $\pt(\mu+b) > 20\gev$, are
\begin{eqnarray}
\sigma({\rm top})[7\tev] &=& \!239\pm53\stat\pm33\syst\pm24\,({\rm theory})\fb\,,  \nonumber \\
\sigma({\rm top})[8\tev] &=& \!289\pm43\stat\pm40\syst\pm29\,({\rm theory})\fb\,. \nonumber 
\end{eqnarray}
The systematic uncertainties are nearly 100\% correlated between the two measurements. 
 
In summary, top quark production is observed for the first time in the forward region.  The cross-section results are in agreement with the SM predictions of $180_{-41}^{+51}(312_{-68}^{+83})\fb$ at $7(8)\tev$ obtained at NLO using MCFM.  The differential distributions of the yield and charge asymmetry are also consistent with SM predictions.

\section*{Acknowledgments}

\noindent We express our gratitude to our colleagues in the CERN
accelerator departments for the excellent performance of the LHC. We
thank the technical and administrative staff at the LHCb
institutes. We acknowledge support from CERN and from the national
agencies: CAPES, CNPq, FAPERJ and FINEP (Brazil); NSFC (China);
CNRS/IN2P3 (France); BMBF, DFG, HGF and MPG (Germany); INFN (Italy); 
FOM and NWO (The Netherlands); MNiSW and NCN (Poland); MEN/IFA (Romania); 
MinES and FANO (Russia); MinECo (Spain); SNSF and SER (Switzerland); 
NASU (Ukraine); STFC (United Kingdom); NSF (USA).
The Tier1 computing centres are supported by IN2P3 (France), KIT and BMBF 
(Germany), INFN (Italy), NWO and SURF (The Netherlands), PIC (Spain), GridPP 
(United Kingdom).
We are indebted to the communities behind the multiple open 
source software packages on which we depend. We are also thankful for the 
computing resources and the access to software R\&D tools provided by Yandex LLC (Russia).
Individual groups or members have received support from 
EPLANET, Marie Sk\l{}odowska-Curie Actions and ERC (European Union), 
Conseil g\'{e}n\'{e}ral de Haute-Savoie, Labex ENIGMASS and OCEVU, 
R\'{e}gion Auvergne (France), RFBR (Russia), XuntaGal and GENCAT (Spain), Royal Society and Royal
Commission for the Exhibition of 1851 (United Kingdom).

\addcontentsline{toc}{section}{References}
\setboolean{inbibliography}{true}
\bibliographystyle{LHCb}
\bibliography{top_paper.bbl}

\newpage



\section*{Supplemental Material}

\begin{figure}[h!]
  \begin{center}
    \includegraphics[width=0.49\textwidth]{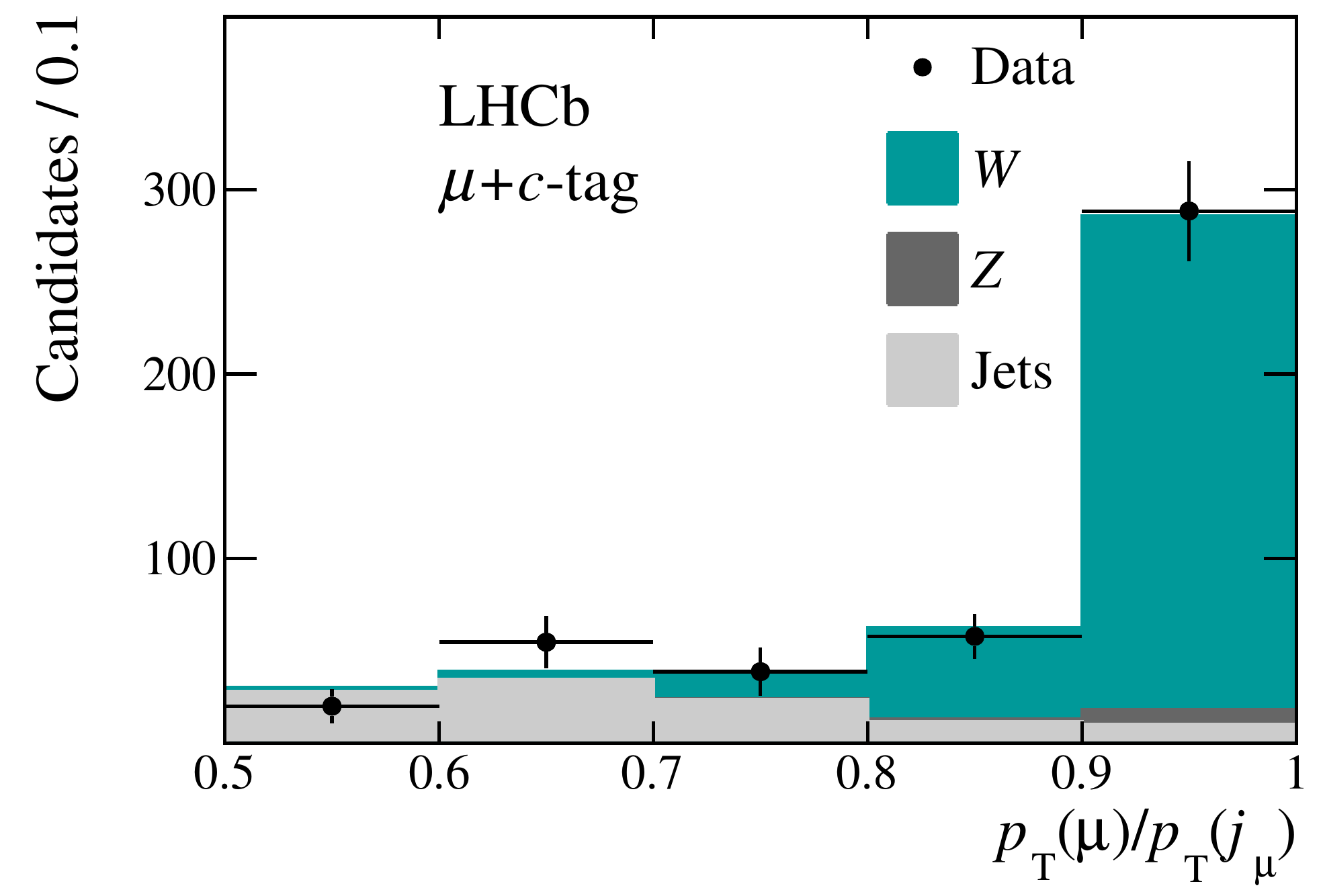}
    \caption{Distribution of $p_{\rm T}(\mu)/p_{\rm T}(j_{\mu})$ with fit overlaid for all $W\!+\!c$ candidates.}
  \end{center}
\end{figure}


\newpage


\centerline{\large\bf LHCb collaboration}
\begin{flushleft}
\small
R.~Aaij$^{38}$, 
B.~Adeva$^{37}$, 
M.~Adinolfi$^{46}$, 
A.~Affolder$^{52}$, 
Z.~Ajaltouni$^{5}$, 
S.~Akar$^{6}$, 
J.~Albrecht$^{9}$, 
F.~Alessio$^{38}$, 
M.~Alexander$^{51}$, 
S.~Ali$^{41}$
% ...

{\footnotesize \it
% ...
$ ^{5}$Clermont Universit\'{e}, Universit\'{e} Blaise Pascal, CNRS/IN2P3, LPC, Clermont-Ferrand, France\\
$ ^{6}$CPPM, Aix-Marseille Universit\'{e}, CNRS/IN2P3, Marseille, France\\
% ...
$ ^{9}$Fakult\"{a}t Physik, Technische Universit\"{a}t Dortmund, Dortmund, Germany\\
% ...
$ ^{37}$Universidad de Santiago de Compostela, Santiago de Compostela, Spain\\
$ ^{38}$European Organization for Nuclear Research (CERN), Geneva, Switzerland\\
% ...
$ ^{41}$Nikhef National Institute for Subatomic Physics, Amsterdam, The Netherlands\\
% ...
$ ^{46}$H.H. Wills Physics Laboratory, University of Bristol, Bristol, United Kingdom\\
% ...
$ ^{51}$School of Physics and Astronomy, University of Glasgow, Glasgow, United Kingdom\\
$ ^{52}$Oliver Lodge Laboratory, University of Liverpool, Liverpool, United Kingdom\\
% ...
}
\end{flushleft}

\end{document}